\documentclass[5p,authoryear]{elsarticle}
\usepackage{lineno}
\modulolinenumbers[5]
\usepackage{color}
\usepackage{colortbl}
\usepackage{amssymb}
\usepackage{amsmath}
\usepackage{graphicx}
\usepackage{url}
\usepackage{subfigure}
\usepackage{graphicx}
\usepackage{tabularx}
\usepackage{bm}
\usepackage{algorithm}
\usepackage{algorithmic}

\newcolumntype{L}{>{\raggedright\arraybackslash}X}
\newcolumntype{a}{>{\hsize=0.6\hsize\raggedright\arraybackslash}X}
\newcolumntype{b}{>{\hsize=0.45\hsize\raggedright\arraybackslash}X}
\newcolumntype{n}{>{\hsize=0.32\hsize\raggedright\arraybackslash}X}
\newcolumntype{s}{>{\hsize=0.12\hsize\raggedright\arraybackslash}X}
\newcolumntype{m}{>{\hsize=0.2\hsize\raggedright\arraybackslash}X}

\definecolor{AD}{rgb}{0,0,0}
\definecolor{AM}{rgb}{0.66,0,0}
\definecolor{AV}{rgb}{0,0.33,0}
\definecolor{CeM}{rgb}{0.66,0.33,0}
\definecolor{CL}{rgb}{0,0.66,0}
\definecolor{CM}{rgb}{0.66,0.66,0}
\definecolor{H}{rgb}{0,1,0}
\definecolor{LD}{rgb}{0.66,1,0}
\definecolor{LGN}{rgb}{0,0,0.5}
\definecolor{LP}{rgb}{0,0.33,0.5}
\definecolor{L-SG}{rgb}{0.66,0.33,0.5}
\definecolor{MDl}{rgb}{0,0.66,0.5}
\definecolor{MDm}{rgb}{0.66,0.66,0.5}
\definecolor{MDv}{rgb}{0,1,0.5}
\definecolor{MGN}{rgb}{0.66,1,0.5}
\definecolor{MV-re}{rgb}{0,0,1}
\definecolor{Pc}{rgb}{0.66,0,1}
\definecolor{Pf}{rgb}{0,0.33,1}
\definecolor{Pt}{rgb}{0.66,0.33,1}
\definecolor{PuA}{rgb}{0,0.66,1}
\definecolor{PuI}{rgb}{0.66,0.66,1}
\definecolor{PuL}{rgb}{0,1,1}
\definecolor{PuM}{rgb}{0.66,1,1}
\definecolor{Pv}{rgb}{0.33,0,0}
\definecolor{R}{rgb}{1,0,0}
\definecolor{VA}{rgb}{0.33,0.33,0}
\definecolor{VAmc}{rgb}{1,0.33,0}
\definecolor{VLa}{rgb}{0.33,0.66,0}
\definecolor{VLp}{rgb}{1,0.66,0}
\definecolor{VM}{rgb}{0.33,1,0}
\definecolor{VMb}{rgb}{1,1,0}
\definecolor{VPI}{rgb}{0.33,0,1}
\definecolor{VPL}{rgb}{1,0,1}
\definecolor{VPLa}{rgb}{1,0,1}
\definecolor{VPLp}{rgb}{1,0,1}
\graphicspath{{./pdf/}{./jpg/}{./png/}}
\DeclareGraphicsExtensions{.pdf,.jpg,.png}

\journal{NeuroImage}

\begin{document}

\begin{frontmatter}

\title{A probabilistic atlas of the human thalamic nuclei combining \emph{ex vivo} MRI and histology}

\author[ucl,bcbl]{Juan Eugenio Iglesias\corref{mycorrespondingauthor}}	 
\cortext[mycorrespondingauthor]{Corresponding author}
\ead{e.iglesias@ucl.ac.uk}
\author[uclm]{Ricardo Insausti}
\author[bcbl]{Garikoitz Lerma-Usabiaga}
\author[drc]{Martina Bocchetta}
\author[martinos,dtu]{Koen Van Leemput}
\author[martinos]{Douglas N. Greve}
\author[martinos]{Andre van der Kouwe}
\author{the Alzheimer's Disease Neuroimaging Initiative\fnref{adniBlob}}
\author[martinos,csail]{Bruce Fischl}
\author[bcbl]{C\'esar Caballero-Gaudes}
\author[bcbl]{Pedro M. Paz-Alonso}
%\author{for the Alzheimer's Disease Neuroimaging Initiative\fnref{adniBlob}}
\fntext[adniBlob]{Data used in preparation of this article were obtained from the Alzheimer's Disease Neuroimaging Initiative (ADNI) database (\url{http://adni.loni.usc.edu}). As such, the investigators within the ADNI contributed to the design and implementation of ADNI and/or provided data but did not participate in analysis or writing of this report. A complete listing of ADNI investigators can be found at: \url{adni.loni.usc.edu/wp-
content/uploads/how_to_apply/ADNI_Acknowledgement_List.pdf}.}

\address[ucl]{Centre for Medical Image Computing (CMIC), Department of Medical Physics and Biomedical Engineering, University College London, United Kingdom}
\address[bcbl]{BCBL. Basque Center on Cognition, Brain and Language, Spain}
\address[uclm]{Human Neuroanatomy Laboratory, University of Castilla-La Mancha, Spain}
\address[drc]{Dementia Research Centre, Department of Neurodegenerative Disease, Institute of Neurology, University College London, United Kingdom}
\address[martinos]{Martinos Center for Biomedical Imaging, Massachusetts General Hospital and Harvard Medical School, USA}
\address[dtu]{Department of Applied Mathematics and Computer Science, Technical University of Denmark}
\address[csail]{MIT Computer Science and Artificial Intelligence Laboratory, USA}

\begin{abstract}

The human thalamus is a brain structure that comprises numerous, highly specific nuclei. Since these nuclei are known to have different functions and to be connected to different areas of the cerebral cortex, it is of great interest for the neuroimaging community to study their volume, shape and connectivity \emph{in vivo} with  MRI. In this study, we present a probabilistic atlas of the thalamic nuclei built using \emph{ex vivo} brain MRI scans and histological data, as well as the application of the atlas to \emph{in vivo} MRI segmentation. The atlas was built using manual delineation of 26 thalamic nuclei on the serial histology of 12 whole thalami from six autopsy samples, combined with manual segmentations of the whole thalamus and surrounding structures (caudate, putamen, hippocampus, etc.) made on \emph{in vivo} brain MR data from 39 subjects. The 3D structure of the histological data and corresponding manual segmentations was recovered using the \emph{ex vivo} MRI as reference frame, and stacks of blockface photographs acquired during the sectioning as intermediate target. The atlas, which was encoded as an adaptive tetrahedral mesh, shows a good agreement with with previous histological studies of the thalamus in terms of volumes of representative nuclei. When applied to segmentation of \emph{in vivo} scans using Bayesian inference, the atlas shows excellent test-retest reliability, robustness to changes in input MRI contrast, and  ability to detect differential thalamic effects in subjects with Alzheimer's disease. The probabilistic atlas and companion segmentation tool are publicly available as part of the neuroimaging package FreeSurfer.

\end{abstract}


\end{frontmatter}

% \linenumbers

\section*{Introduction}

\noindent
The thalamus is a diencephalic structure located between the cortex and the midbrain. Traditionally, the thalamus has been considered primarily a link in the flow of sensory signals, through its white matter connections to virtually the entire cortex \citep{johansen2005functional}. However, current views suggest that the thalamus is more than a simple ``relay station'', and continues to contribute to the processing of information within cortical hierarchies \citep{sherman2007thalamus,sherman2016thalamus}.
Among other functions, the thalamus is involved in  the regulation of consciousness, sleep and alert states; the motor system; and spoken language \citep{sherman2001exploring}. The study of these functions  with MRI has attracted wide attention from the neuroimaging community \citep{fernandez2010reductions,czisch2004functional,guye2003combined,binder1997human}, and so has the \emph{in vivo} study of pathologies associated with the thalamus, such as schizophrenia \citep{buchsbaum1996pet,andreasen1994thalamic}, Alzheimer's disease \citep{de2008strongly,zarei2010combining}, epilepsy \citep{natsume2003mri,bonilha2005voxel}, Huntington's disease \citep{aron2003inhibition,kassubek2005thalamic} or dyslexia \citep{diaz2012dysfunction,giraldo2015morphological,jednorog2015reliable}.  

Segmentation of the whole thalamus in structural MRI is a prerequisite for most MRI-based studies of this structure, and many methods have been developed to produce automated segmentations. \citet{fischl2002whole} used a voxel-wise probabilistic atlas of anatomy and MRI intensities to segment the thalamus, along with a number of other brain structures; this method is implemented in the widespread, open-source package FreeSurfer \citep{fischl2012freesurfer}.
\citet{patenaude2011bayesian} used a combined model of shape and appearance, also to segment a set of brain structures including the thalamus; an implementation of this method (``FIRST'') is available as part of the popular FSL package \citep{smith2004advances}. A number of standard segmentation algorithms have also been applied to thalamus segmentation in structural MRI, such as multi-atlas segmentation \citep{heckemann2006automatic}, fuzzy clustering \citep{amini2004automatic}, voxel classification \citep{zikic2013atlas} or Bayesian segmentation \citep{puonti2016fast}. 

However, the thalamus is not a homogeneous structure; it consists of several nuclear masses, which serve multiple functions. While many studies agree on the existence of 14 major nuclei, these can be subdivided histologically into many more subnuclei, so the exact number depends on the level of detail of the classification \citep{jones2012thalamus,morel2007stereotactic,mai2012thalamus}. \emph{In vivo} segmentation of these nuclei in MRI can enable neuroimaging studies of the thalamus (e.g., morphometry; structural and functional connectivity) at a much higher level of specificity, and also has the potential to provide more accurate surgical planning and more precise placement of deep brain stimulation (DBS) devices. These applications have sparked the interest of the neuroimaging community in automated segmentation algorithms for the thalamic nuclei.

Many thalamic nuclei segmentation methods  have  been based on applying  clustering techniques to diffusion MRI data, which provide more contrast between the nuclei than structural MRI scans -- despite their lower resolution. A subset of these techniques have focused on the local diffusion properties of the tissue. For example, \citet{mang2012thalamus} clustered the voxels inside the thalamus into 21 predefined groups using the direction of the leading eigenvector of the diffusion tensor.
\citet{duan2007thalamus} used the mean shift algorithm \citep{fukunaga1975estimation} with the Frobenius distance between the tensors.
\citet{wiegell2003automatic} used the k-means algorithm to create 14 clusters, using a distance function that was a linear combination of the Mahalanobis voxel distance and the Frobenius tensor distance. 
\citet{jonasson2007level} also used k-means, but only to initialize surfaces that then evolved with the level set method; the cluster prototypes were given by the tensors minimizing the variation within the groups.
% -- rather than the mean tensors. 
Recently, \citet{battistella2017robust} moved away from the tensor model and used a spherical harmonic representation of the full orientation distribution function, to cluster the voxels with k-means.  

Local diffusion information is typically insufficient to discriminate between thalamic nuclei. In their pioneering work, \citet{behrens2003non} and \citet{johansen2005functional} used probabilistic tractography to parcellate the thalamus according to the connectivity of its voxels with predefined target regions of the cerebral cortex. This approach and its extensions (e.g., \citealt{kasenburg2016structural,abivardi2017deconstructing}) have been used in multiple applications, such as the definition of targets in DBS \citep{akram2018connectivity,middlebrooks2018method}. Other  approaches have relied on clustering connectivity patterns to parcellate the thalamus; the connectivity can be structural (i.e., derived from diffusion MRI), as in \citet{lambert2017defining}, or functional (i.e., estimated with resting state functional MRI), as in \citet{ji2016dynamic,hale2015comparison}.

Other thalamic parcellation efforts have relied on supervised machine learning techniques. For example, \citet{stough2014automatic} used local measures (fractional anisotropy, eigenvector directions) and connectivity with cortical regions as input features to a random forest \citep{breiman2001random} in order to divide the thalamus into six major nuclei. Supervision beyond the use of cortical regions for connectivity-based parcellation ameliorates the main disadvantage of unsupervised  methods, which is the lack of correspondence of the clusters with the underlying anatomy -- even if an expert attempts to manually map clusters to nuclei, as in \citet{wiegell2003automatic}.  

Creating ground truth segmentations on \emph{in vivo} scans can be difficult, particularly when one tries to segment thalamic nuclei defined at finer levels of division -- when the resolution (especially in diffusion MRI) and contrast are insufficient. For this reason, some works have used histology  to create atlases of the thalamus. \citet{krauth2010mean} used manual segmentations on the histology of six cases \citep{morel2007stereotactic} in order to compute an average thalamus with 42 nuclei. In follow-up work, \citet{jakab2012generation} mapped their average to MNI space to be able to use it in registration-based segmentation. \citet{sadikot2011creation} used a single labeled case, which they also mapped to a reference space (Colin27, \citealt{holmes1998enhancement}) to be able to segment new cases. These approaches inherit the limitations of registration-based segmentation: the inability of a single subject-segmentation pair to cover the spectrum of variability of a larger population, and the inability to accurately register across MRI contrasts.

In this work, we present a probabilistic atlas of the human thalamus and its nuclei, as well as surrounding anatomy (the latter is important to enable segmentation of thalamus using Bayesian inference). The atlas was derived from manual segmentations of the thalamic nuclei on the histological images of 12 whole thalami from six autopsy samples, as well as delineations of the whole thalamus and surrounding structures (e.g., caudate, putamen, hippocampus) in 39 T1 scans acquired at standard resolution (i.e., $\sim$ 1 mm). The 3D reconstruction of the histology was assisted by \emph{ex vivo} MRI scans and blockface photographs. Compared with previous histology-based atlases of the thalamus \citep{krauth2010mean,sadikot2011creation}, our proposed atlas is probabilistic, models surrounding anatomy, and can be used in combination with Bayesian inference in order to directly segment MRI scans of arbitrary contrast.
% without going through an intermediate reference space.

\section*{Materials and methods}

\subsection*{Ex vivo specimens}

\noindent
In order to build the proposed atlas, we used data from six  \emph{post mortem} cases from the body donor program of the University of Castilla - La Mancha Medical School (Albacete, Spain). The demographic data of the cases is shown in Table~\ref{tab:demographics}. The brain fixation was performed  \emph{in situ} by personnel of the Human Neuroanatomy Laboratory, by neck disection of both primitive carotids in the lower third of the neck, followd by cannulation of the carotids. The fixation started with a flush of 4 l of saline, followed by 8 l of 4\% paraformaldehyde in phosphate buffer (pH 7.4). In order to allow the fixative to flow, the internal jugular vein was sectioned on one side. After perfusion, the brain was left \emph{in situ} for 48 hours, and subsequently extracted following standard autopsy procedures. Postfixation until scanning was carried out by storage in a container filled with 4\% paraformaldehyde. This \emph{in situ} fixation method better preserves the shape of the individual brain, fitting exactly the intracranial shape (as opposed to a generic container), and minimizes the impact of the extraction procedure on the probabilistic atlas to be built.

\begin{table}[t]
\centering
\begin{tabularx}{\linewidth}{| b | n | n | b | b |}
\hline
Case & Age at \newline death & Gender & Brain \newline weight  & PMI\\
\hline
HNL4\_13 & 97 & male & 1.238 Kg & 9h \\ 
HNL7\_14 & 98 & female & 1.168 Kg & 6h \\
HNL5\_13 & 59 & male & 1.020 Kg & N/A \\ 
HNL8\_14 & 61  & female & 1.409 Kg & 10h  \\ 
HNL14\_15 & 87 & male & 1.100 Kg & 2h 30m \\ 
HNL16\_16 & 84 & male & 1.264 Kg & 3h 30m\\ 
\hline
\end{tabularx}
\caption{Demographics of the \emph{ex vivo} cases that were used to build the atlas. PMI stands for ``post mortem interval''.}
\label{tab:demographics}
\end{table}

\subsection*{MRI scanning and processing}

\noindent
\emph{Ex vivo} MR images of the whole brains were acquired on a 3 T Siemens Magnetom TIM Trio scanner with a 12 channel receiver coil; despite the reduced efficiency compared with the 32 channel coil, it enables acquisition at higher resolution without running out of RAM in the image reconstruction. The brains were scanned in vacuum bags filled with Fluorinert FC-3283 (3M, Maplewood‎, MN, U.S.A.), in order to minimize the negative impact of air bubbles and susceptibility artifacts. The images were acquired with a 3D multi-slab balanced steady-state free precession sequence \citep{mcnab2009high} with TE/TR = 5.3/10.6 ms and flip angle $35^\circ$. Four axial slabs with 112 slices each were used to cover the whole volume of the brains, and 57\% slice oversampling was used in order to minimize slab aliasing. The resolution of the scans was 0.25 mm isotropic, with matrix size 720$\times$720$\times$448 voxels (axial). MR images were acquired with four different RF phase increments (0, 90, 180, 270 degrees) and averaged to reduce banding artifacts. The time of acquisition per phase was 90 minutes. Ten repetitions of this protocol were acquired for increased signal-to-noise ratio (SNR). The total length of the protocol was thus 60 hours. 

\begin{figure}[t!]
\centering
\includegraphics[width=.5\textwidth]{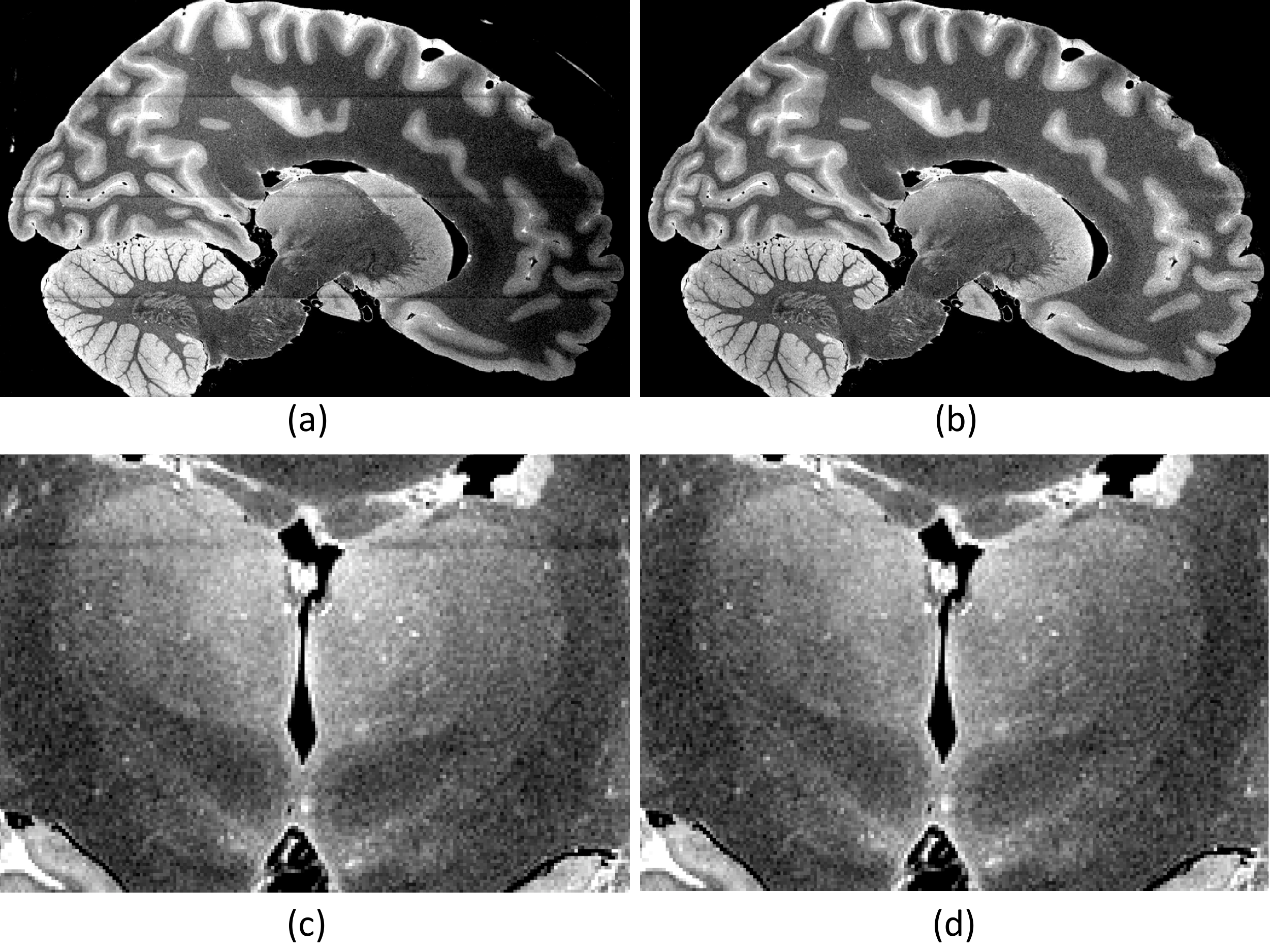}
\caption{(a)~Sample sagittal slice of \emph{ex vivo} MRI scan of case NHL8\_14. (b)~Corrected for bias field and slab boundary artifacts. (c)~Close-up of left thalami in coronal view, uncorrected. (d)~Corrected version of (c).}
\label{fig:sampleMRI}
\end{figure}

Combined with the 12 channel receiver coil, the multi-slab acquisition described above enabled us to bypass the memory limitations of our clinical scanner when reconstructing the images, while preserving the SNR efficiency of 3D acquisitions. However, this type of acquisition also introduces slab boundary artifacts at the interfaces between the slabs. After computing a brain mask with simple Otsu thresholding \citep{otsu1975threshold}, such artifacts were corrected simultaneously with the bias field using a Bayesian method \citep{iglesias2016simultaneous}. Sample slices of the MRI scans are shown in Figure~\ref{fig:sampleMRI}.

\subsection*{Histological analysis}

\noindent
After MRI scanning, the specimens were prepared for histological analysis. First, the brains were separated into left and right hemispheres.
Then, a brain sectioning knife was used to make a transverse cut perpendicular to the intercommissural line. 
Using this plane of section, parallel blocks (thickness 10-14 mm) were extracted, from the frontal to the occipital pole. 
For each case and hemisphere, and in order to avoid incomplete sectioning of the whole thalamus, the anterior limit of the first block was sectioned at the level of the anterior hypothalamus and head of the caudate nucleus. The blocks containing thalamic tissue (three or four, depending on the case) were selected for histological analysis, numbered and subsequently immersed in cryoprotectant (Glicerol 10\% and subsequently 20\%) during 120 hours at $4^\circ$C temperature. After this procedure, the tissue blocks were sectioned at 50 micron intervals using a sliding microtome, which was coupled to a freezing unit and covered in dry ice. A blockface photograph was taken before each cut using an Olympus E-420 camera mounted on a shelf over the microtome; these photographs are  useful for post-hoc 3D reconstruction. One every ten sections (i.e., every 0.5 mm) was selected for Nissl staining with thionin and posterior cytoarchitectonic analysis and segmentation. The other nine were preserved for future analyses with complementary techniques. The selected, Nissl-stained sections were digitized at 4 micron resolution using an Epson Perfection V800 Photo flatbed scanner. A corresponding pair of blockface and histological images are shown in Figure~\ref{fig:blockfaceHisto}.

\subsection*{Blockface photograph processing}

\begin{figure}[t!]
\centering
\includegraphics[width=.4\textwidth]{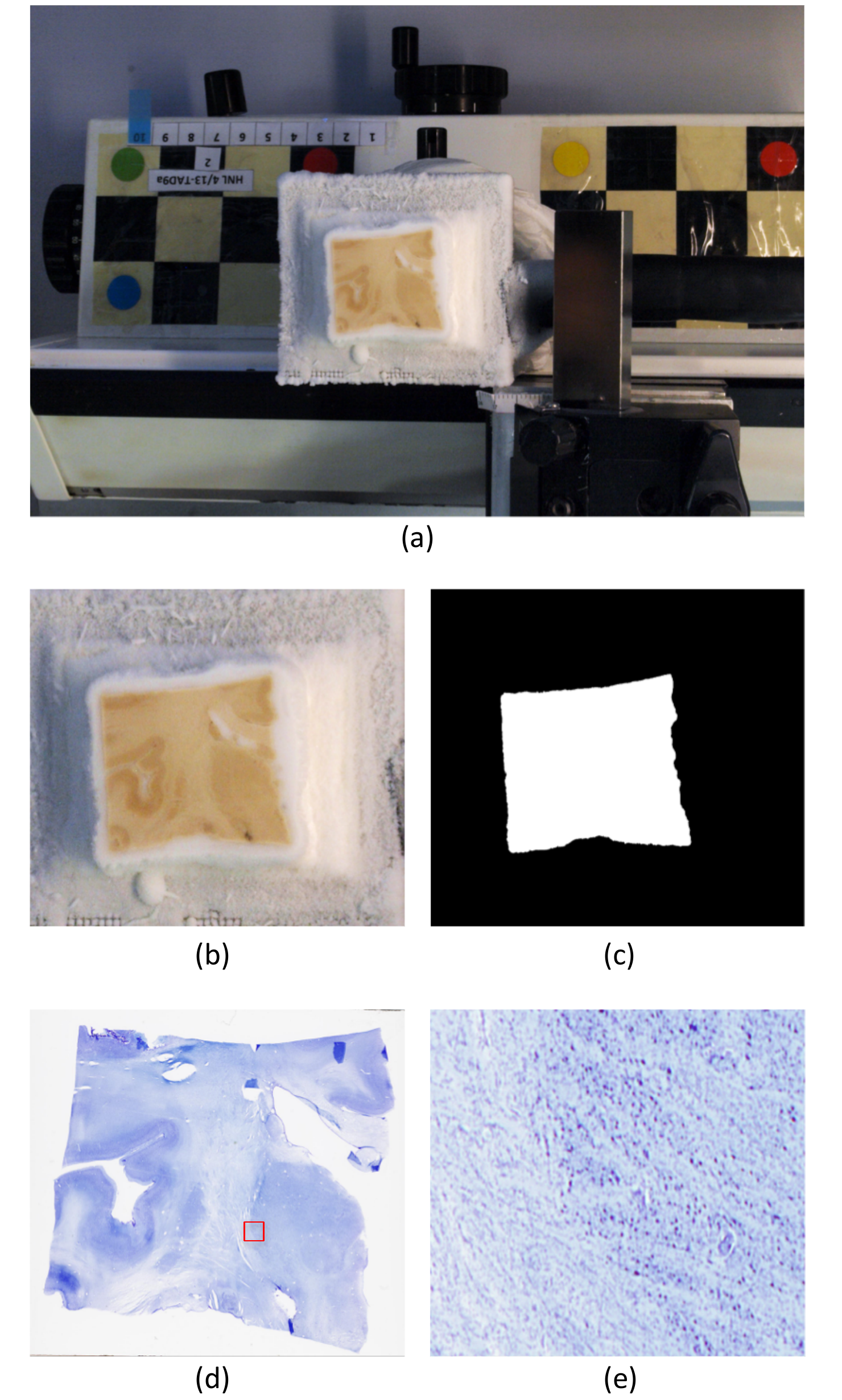}
\caption{Histology processing: (a)~Sample blockface photograph of case HNL4\_13. (b)~Homography corrected photograph. (c)~Automated segmentation with random forest classifier. (d)~Corresponding digitized histology. (e)~Close up of the region inside the red square in (d).}
\label{fig:blockfaceHisto}
\end{figure}

\noindent
In order to be useful in histology reconstruction, the blockface images need to be registered and perspective corrected, such that, when stacked, they render a volume that is 3D consistent. Registration is needed to correct for the perspective distortion introduced by small movements of the microtome and the camera setup. In order to co-register the images, we used as reference a photograph of the microtome without any sample on the block holder. On this photograph, we manually marked the corners of the block holder, and used them to define a binary mask covering the whole image domain except for the block holder -- so that the registration is not influenced by the brain samples.  The registration was performed by detecting salient points inside the mask with SURF \citep{bay2006surf}, matching their (SURF) feature vectors, and robustly optimizing a homography transformation with RANSAC \citep{fischler1987random}. In order to introduce very salient points and thus ease the registration, we glued a checkerboard pattern and round stickers in different colors to the microtome (see Figure~\ref{fig:blockfaceHisto}a).

After registration, we used an homography transform to correct for scaling and geometric perspective distortion. The homography was computed by matching the  four corners of the block holder in the reference image to a rectangular grid of size equal to the holder's physical dimensions, defined at 0.1 mm resolution. The transform was precomputed using the reference photograph, and then applied to all the other blockface images in order to extract perspective corrected images of known resolution. A sample corrected image is shown in Figure~\ref{fig:blockfaceHisto}b.

Finally, we segmented the tissue from the underlying block holder (which was surrounded by dry ice) using a random forest pixel classifier based on visual features \citep{geremia2011spatial,criminisi2013decision}. The classifier was trained on 12 pseudorandomly selected, manually labeled photographs -- one from each case and side. We found this small training dataset to perform sufficiently well, due to the large contrast between the tissue and the underlying dry ice (see example if Figure~\ref{fig:blockfaceHisto}c).

\subsection*{3D reconstruction of histology via blockface photographs}

\begin{figure*}[t!]
\centering
\includegraphics[width=0.8\textwidth]{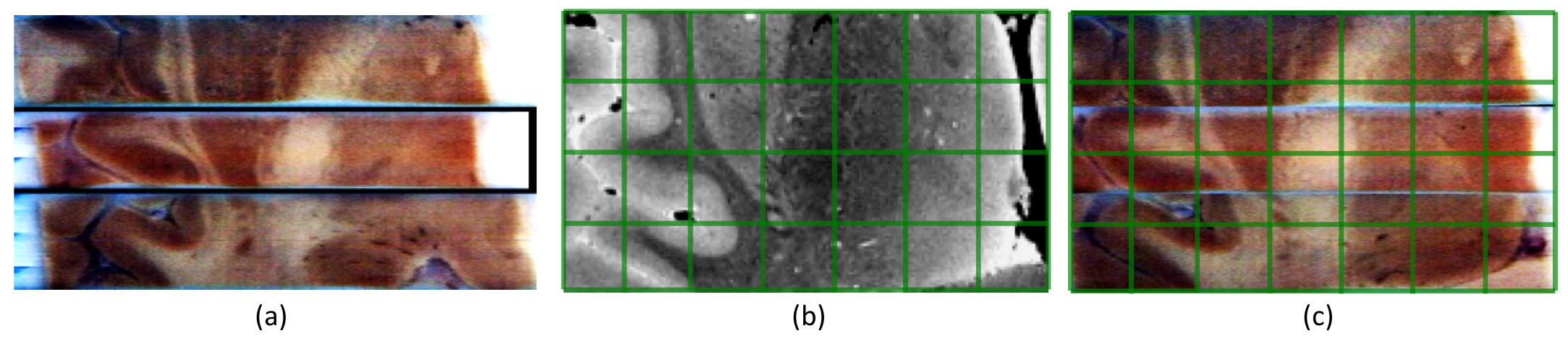}
\caption{Rigid alignment of stacks of blockface photographs to MRI. (a)~Initialization: axial view across the thalamus. (b)~MRI. (c)~Aligned stacks. We have overlaid a grid on subfigures (b) and (c) to ease the visual assessment of the quality of the registration.}
\label{fig:rigidAlignment}
\end{figure*}

\noindent
Recovering the 3D structure of the histology requires first stacking the sections corresponding to each block (which we know are parallel, and with what separation), and subsequently estimating two sets of transforms: 1. nonrigid registrations of each section within a block, in order to correct for the tissue shrinkage, deformation and occasional tear and folding that occur when the sections are stained and mounted on glass slides; and 2. rigid registrations that align the blocks with each other. Computing these transforms -- particularly the nonlinear -- using only histological data is an  ill-posed problem; without any additional information, we can only resort to registering each section with its neighbors, which is well known to cause geometric distortions such as the ``banana effect'' (straightening of curved structures) and ``z-shift'' (accumulation of errors along the stack). 

A better alternative is to guide the reconstruction with 3D consistent data acquired with another modality, typically MRI. However, using the MRI volume directly is still complicated, since solving for the pose of the blocks and the nonlinear registrations simultaneously is also ill posed \citep{malandain2004fusion}. More precise solutions can be achieved by using 3D consistent images of the whole blocks as intermediate target. These images play the role of stepping stones between the histology and the MRI, since they can be linearly registered to the MRI to obtain the pose of the blocks. 
For example, \citet{adler2018characterizing} used MRI scans of the blocks as intermediate data. This approach makes the registration to the original MRI scan of the whole brain easier, since it is an intra-modality problem. However, it has the disadvantage that it still requires estimating a rigid transform between the MRI scan of the block and the histology stack -- albeit much easier to estimate than that between the MRI of the whole brain and the histology. Here, we followed \citet{amunts2013bigbrain} instead, and used the stacks of (perspective corrected, segmented) blockface photographs. This choice has the advantage that the exact correspondence between the histological sections and photographs in the stack is known. On the other hand, the registration between the intermediate images and the original MRI is slightly harder because it is inter-modality, but, since it is a rigid registration problem, mutual information works well.

More specifically, we first rigidly aligned the stacks of photographs to the whole brain MRI, using the segmentations produced by the Otsu thresholding (MRI) and the pixel classifier (photographs) to mask the cost function, which used mutual information. We used an iterative algorithm, in which we alternately updated a global registration of the whole brain MRI to the set of blocks, and then refined the set of individual block transforms that aligned them with the MRI. The algorithm was initialized by stacking the blocks with a 1 mm spacing between them (which approximates the tissue that is lost when cutting the blocks), and aligning them with 2D rigid transforms and mutual information. 
 The pose of the whole brain MRI was initialized by coarse manual alignment. The rigid co-registration algorithm is illustrated with an example in Figure~\ref{fig:rigidAlignment}.

\begin{figure*}[t!]
\centering
\includegraphics[width=0.85\textwidth]{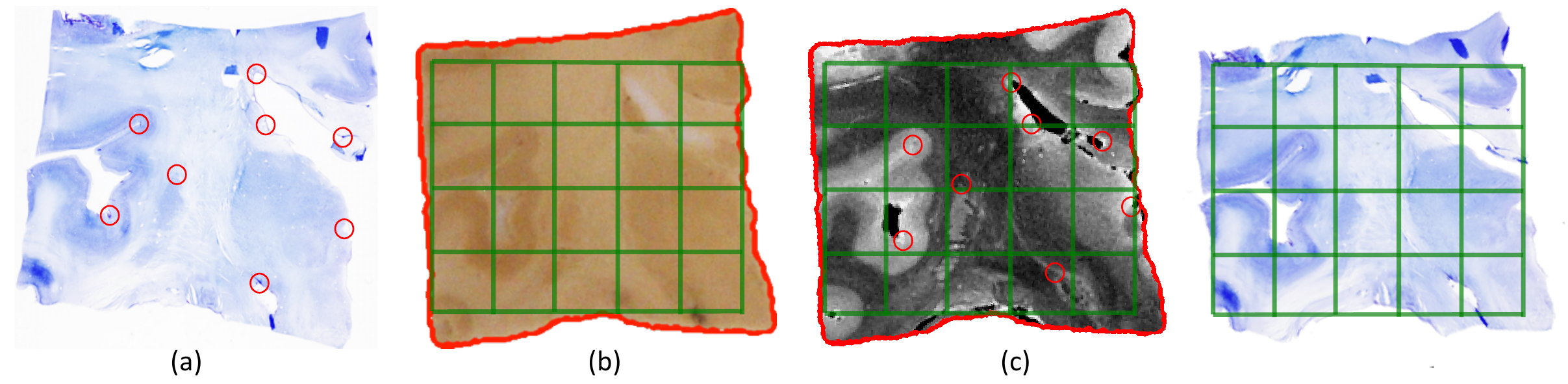}
\caption{Nonrigid registration of histology and MRI. (a)~Sample histological section, same as in Figure~\ref{fig:blockfaceHisto}. (b)~Corresponding blockface photograph, masked by the corresponding automated segmentation. (c)~MRI scan resampled to the space of the photograph. (d)~Registered histology. The manually placed landmarks are marked with red circles in subfigures (a) and (c).}
\label{fig:nonrigidAlignment}
\end{figure*}

\begin{figure}[t!]
\centering
\includegraphics[width=0.5\textwidth]{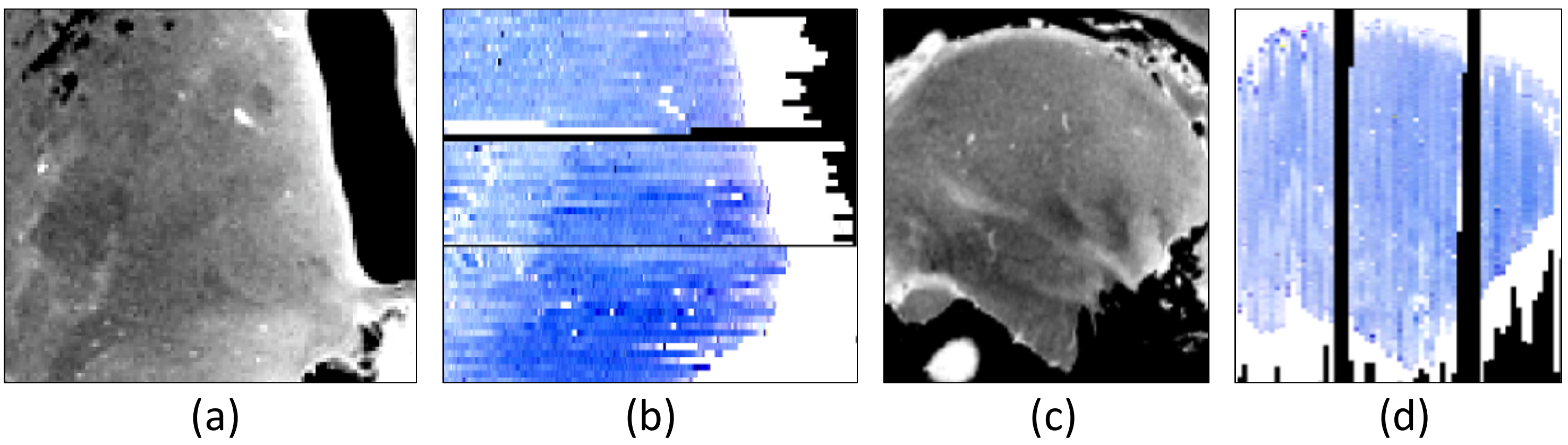}
\caption{Examples of reconstructed histology. (a)~Axial view of thalamus in MRI. (b)~Corresponding slice through the stack of reconstructed histology. (c)~Sagittal slice of thalamus in MRI. (d)~Corresponding histology. }
\label{fig:reconHisto}
\end{figure}

Given the pose of the blocks, it is still necessary to compute the nonlinear registration of each stained section. For this purpose, we first took the photograph to which the section corresponded, resampled the MRI volume onto it with the corresponding rigid transform, and masked it with the corresponding mask, given by the pixel classifier. This resampled MRI was used as target to register the corresponding histological section. We used Elastix \citep{klein2010elastix} for the registration, combining mutual information with a B-spline transform (control point spacing: 3 mm). We found the registration based solely on image intensities not to be robust enough for our application, particularly when tears and folds had occurred when mounting the tissue on the slide. To increase the robustness, we manually placed pairs of corresponding landmarks on the images (between 6 and 18 per pair of images). The cost function was the sum of the mutual information and the mean distance between corresponding landmarks after registration. The nonrigid registration  is illustrated with an example in Figure~\ref{fig:nonrigidAlignment}. Sample orthogonal slices of the reconstructed histology stacks are shown in Figure~\ref{fig:reconHisto}.

\begin{table*}[!h]
\scriptsize
\centering
\begin{tabularx}{\linewidth}{| l | l | l | m  | a |}
\hline
Group & Color & Abbrev. & Nucleus & Definition\\
\hline
Anterior &\cellcolor{AV} & AV & Anteroventral & Well defined nucleus, starting very rostrally. Continued by the LD. Small/medium sized neurons. We include the anterior medial and anterior dorsal nuclei into the AV.  \\
\hline

Lateral&\cellcolor{LD} & LD & Laterodorsal & Made up of small cells, pale and homogeneously distributed.  \\
\hline
&\cellcolor{LP} & LP & Lateral  posterior & Group of loosely arranged small and medium neurons. It continues the ventral lateral nucleus an its posterior part (VLp) caudally, as far as the PuA. \\
\hline

Ventral &\cellcolor{VA} & VA & Ventral  anterior & Located at the anterior pole of the thalamus, and formed by medium size neurons crossed by bundles of fibers. \\
\hline
&\cellcolor{VAmc} & VAmc & Ventral anterior  magnocellular &  Formed by big and dark neurons, loosely arranged.   \\
\hline
&\cellcolor{VLa} & VLa &  Ventral lateral   anterior & Formed by small neurons in clusters, in the dorsolateral part of the nucleus. 
 \\
\hline
&\cellcolor{VLp} & VLp & Ventral lateral   posterior & Made up of large neurons, loose appearance.  \\
\hline
&\cellcolor{VPL} & VPL & Ventral  posterolateral & Formed by small and medium sized neurons from the ventral part of VLp to the PuI and PuL. The medial portion (ventral posteromedial nucleus) is included in our definition of VPL. \\
\hline
&\cellcolor{VM} & VM & Ventromedial & The neurons are similar to VA neurons, but without bundles of crossing fibers. It lies ventral to VA. \\
\hline

Intralaminar &\cellcolor{CeM} &  CeM & Central  medial & Formed by a compact group of dark neurons, located close to MV-Re and Pv. \\
\hline
&\cellcolor{CL} & CL & Central  lateral & Made up of big neurons, arranged in clusters. It lies dorsal to the MD, lateral to MDl, and underneath AV and LD. \\
\hline
&\cellcolor{Pc} & Pc & Paracentral & Lateral to MDl. Medial to VLp. Small and connected islands, loose. \\
\hline
&\cellcolor{CM} & CM & Centromedian & Formed by small, condensed neurons. It is surrounded by fibers of the internal medullary lamina. \\
\hline
&\cellcolor{Pf} & Pf & Parafascicular & Formed by small and compact neurons. It lies ventral and medial to the CM.  \\
\hline

Medial &\cellcolor{Pt} & Pt & Paratenial & Rostrocaudally oriented group of small neurons  along the stria medullaris.  \\
\hline
&\cellcolor{MV-re} & MV-re & Reuniens  (medial ventral) & Rostrally situated, it consists of a mix of large and small neurons. Fused with the other side through the adhesion interthalamica. Anteroventral to CM and medial to VA. \\
\hline
&\cellcolor{MDm} & MDm & Mediodorsal medial   magnocellular & Made up of big and darkly stained neurons, sometimes in irregular groups at the ventral part.   \\
\hline
&\cellcolor{MDl} & MDl & Mediodorsal lateral  parvocellular & Smaller neurons which form varied forms of groupings. Bordered by the Pc, CL and Pf ventrally.   \\
\hline

Posterior &\cellcolor{LGN} & LGN & Lateral  geniculate & Formed by magnocellular layers ventrally, and parvocellular dorsally. \\
\hline
&\cellcolor{MGN} & MGN & Medial  Geniculate & Located medial and posterior to the LGN. It is made up of three parts: magnocellular, dorsomedial, and ventromedial.  \\
\hline
&\cellcolor{L-SG} & L-SG & Limitans  (suprageniculate) & Made up of medium and large, deeply stained neurons, which form islands with an irregular profile. It lies on top of the pretectal complex. \\
\hline
&\cellcolor{PuA} & PuA &  Pulvinar anterior & Group of neurons located ventromedially to the LP. \\
\hline
&\cellcolor{PuM} & PuM &   Pulvinar medial & Formed by small and pale neurons of uniform appearance and distribution. It lies at the posterior end of the thalamus. \\
\hline
&\cellcolor{PuL} & PuL &  Pulvinar lateral & Big in size, it occupies most of the lateral part of the caudal thalamus. It is crossed by many fibers.\\
\hline
&\cellcolor{PuI} & PuI &  Pulvinar inferior & Located ventrally and laterally to the PuM, and formed by small and medium neurons. \\
\hline

Others &\cellcolor{R}  & R & Reticular & Groups of medium or large neurons, which wrap the thalamic nuclei. Traversed by numerous bundles of fibers along its extent, and separated from the VA, VLa, VLp, VPL, PuL by the external medullary lamina.  \\

%&\cellcolor{MDv} & MDv & X && Z \\
%&\cellcolor{AD} &  AD & X && Z \\
%&\cellcolor{AM} & AM & X && Z \\
%&\cellcolor{H} & H & X && Z \\
%&\cellcolor{Pv} & Pv & X && Z \\
%&\cellcolor{VMb} & VMb & X && Z \\
%&\cellcolor{VPI} & VPI & X && Z \\
%&\cellcolor{VPLa} & VPLa & X && Z \\
%&\cellcolor{VPLp} & VPLp & X && Z \\
\hline
\end{tabularx}
\caption{Summary of protocol for manual delineation of the thalamic nuclei on the histological images.}
\label{tab:definitionOfNuclei}
\end{table*}

\subsection*{Manual segmentation of nuclei on histology}

\noindent
The analysis of the histological sections was carried out by an expert neuroanatomist (R.I.), using a stereo microscope (Leica EZ4) with 50$\times$ magnification, an optic microscope (Nikon Eclipse E600), and the digitized histology. R.I. manually delineated the nuclei with ITK-Snap (\url{http://www.itksnap.org}), following the rostrocaudal axis of the thalamus, from the anterior thalamus to the end of the pulvinar nuclei. The delineation protocol was based on the characterization of the human and mammalian thalamus by \citet{jones2012thalamus}, and is summarized in Table~\ref{tab:definitionOfNuclei}. Further details on the protocol and criteria for delineation will be described in an additional publication in a specialized journal.

\subsection*{3D reconstruction of manual segmentations: filling the gaps}

\begin{figure}[t!]
\centering
\includegraphics[width=0.5\textwidth]{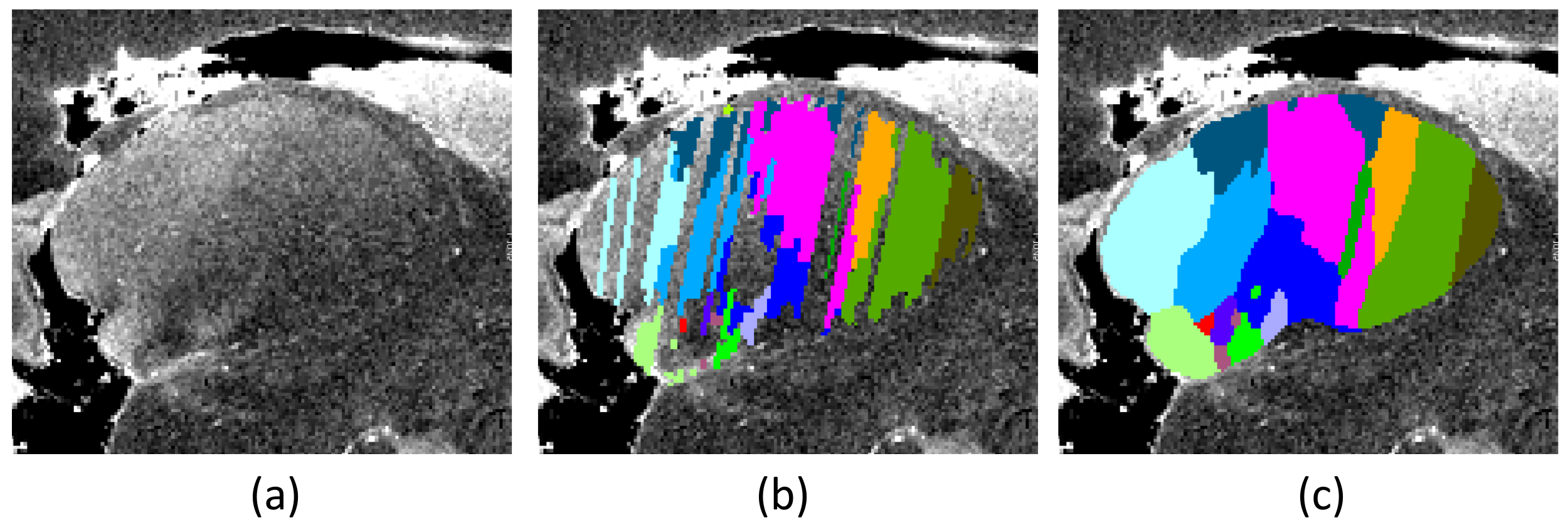}
\caption{Filling the gaps between blocks and refining the segmentation. (a)~Sample sagittal slice of the thalamus. (b)~Propagated manual segmentations. (c)~Labels estimated by minimizing Equation~\ref{eq:histoFixCost}. The color coding is the same as in Table~\ref{tab:definitionOfNuclei}.}
\label{fig:histoFix}
\end{figure}

\noindent
The rigid and nonrigid deformations that were computed to recover the 3D structure of the histological images can be directly applied to the manual segmentations in order to warp them to the space of the MRI scan. However, the directly warped labelings do not immediately yield usable 3D segmentations due to the gaps between blocks and the inconsistencies between the manual segmentations of adjacent sections (see Figure \ref{fig:histoFix}a). To refine the segmentation in MRI space, we used the following automated approach. First, we performed a 2D erosion on each section and label  (including the background) with a small circular kernel  (radius: 1 mm), in order to generate a thin uncertainty zone around the edges of the segmentation -- which models the registration error. Next, we deformed these  eroded segmentations to the space of the MRI scan. Finally, we automatically assigned labels to the eroded voxels, as well as to the voxels in the the gaps between blocks, by minimizing the following cost function:
\begin{align}
\mathcal{C}({\bm{S}})  = & - \sum_j  \log p (I_j; \bm{\theta}_{S_j}) + \alpha \sum_j  D(j; S_j) \ldots \nonumber \\
& + \beta \sum_j  \sum_{j' \in \Gamma(j)} \delta(S_j \neq S_{j'}), 
\label{eq:histoFixCost}
\end{align}
where $\bm{S}=\{S_j\}$ is the segmentation we want to compute ($S_j$ is the segmentation of voxel $j$); $I_j$ represents the MRI intensity at voxel $j$; $\bm{\theta}_l$ are sets of Gaussian parameters associated with each label $l$;  $D(j; S_j)$ represents the (physical) distance of voxel $j$ to the initial segmentation of label $S_j$; $\delta(\cdot)$ is the Kronecker delta; $\Gamma(j)$ is the 6-neighborhood of voxel $j$; and $\alpha$ and $\beta$ are nonnegative weights.

The first term in Equation~\ref{eq:histoFixCost}  encourages voxels with similar intensities to share the same label. The parameter set $\bm{\theta}_l$ correspond to the weights, means and variances of a Gaussian mixture model associated with label $l$. These were estimated using the voxels labeled as $l$ in the initial, eroded segmentation. The second term  penalizes distance from the original segmentation; we set the distance $D(j; S_j)=-\infty$ for voxels that are inside the  segmentation, which effectively preserves these labels in the minimization. The third term is a Markov random field that penalizes pairwise differences in labels of neighboring voxels, thus ensuring the smoothness of the final result. 

In order to minimize the cost in Equation~\ref{eq:histoFixCost}, we used $\alpha$-expansion \citep{boykov2001fast}. 
% Even though convergence to the global minimum is not guaranteed by this algorithm, it yields good results in practice. 
The number of Gaussian components was set to three for the background, and one for all other labels. The relative weights of the terms were set to $\alpha=\beta=1$. A sample output of the algorithm is shown in Figure \ref{fig:histoFix}.

\subsection*{Atlas construction}

\noindent
In order to build a probabilistic atlas of the thalamic nuclei and surrounding tissue, we used our previously presented atlas construction method \citep{van2009encoding,iglesias2015computational}. This method uses Bayesian inference to estimate the probabilistic atlas that most likely generated a training dataset of manual segmentations made on a combination of  \emph{in vivo} and \emph{ex vivo} datasets. By combining segmentations of the whole thalamus and surrounding structures (made on the \emph{in vivo} data) with segmentations of the thalamic nuclei (made on \emph{ex vivo} images), we can derive an atlas that includes both the thalamic nuclei and the surrounding (whole) structures.

\begin{figure*}[t!]
\centering
\includegraphics[width=\textwidth]{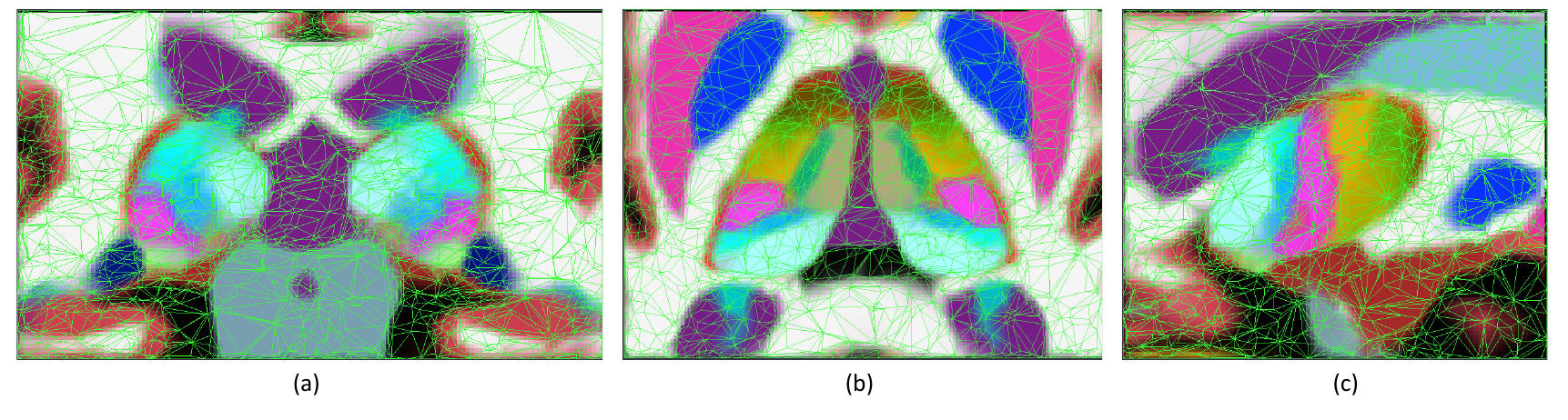}
\caption{Probabilistic atlas, with tetrahedral mesh superimposed. The color of each voxel is a linear combination of the colors in Table~\ref{tab:definitionOfNuclei}, weighted by the corresponding label probabilities. For the surrounding structures, we used the standard FreeSurfer color map. (a)~Sample coronal slice. (b)~Axial slice. (c)~Sagittal slice.}
\label{fig:atlas}
\end{figure*}

In this study, we combined our reconstructions of the thalamic nuclei with manual delineations at the whole structure level made on  39 \emph{in vivo}, T1-weighted scans, acquired at 1 $\times$ 1 $\times$ 1.25 mm resolution (sagittal) on a Siemens 1.5 T plaform with an MP-RAGE sequence. Thirty-six structures, including the left and right whole thalamus, were manually delineated using the protocol described by \citet{caviness1989magnetic}. We note that this is the dataset that was used  to build the atlas in Freesurfer \citep{fischl2002whole,fischl2012freesurfer}; further details on the acquisition can be found in the original publication. Both the \emph{in vivo} and \emph{ex vivo} datasets were augmented with left-right flips to increase their effective size. 
The atlas is represented as a tetrahedral mesh, which is locally adaptive to the complexity of the anatomy \citep{van2009encoding}. Sample slices of the atlas are displayed in Figure~\ref{fig:atlas}. 

\subsection*{Segmentation of in vivo MRI}

\noindent
Given a probabilistic atlas and a generative model of MRI scans, segmentation can be posed as a Bayesian inference problem within the model. As in previous work \citep{van1999automated,iglesias2015computational,iglesias2015bayesian,puonti2016fast,saygin2017high}, and following the literature of Bayesian segmentation \citep{wells1996adaptive,zhang2001segmentation,ashburner2005unified,pohl2006bayesian}, we assumed the following forward model: first, the atlas is spatially warped following a deformation model \citep{ashburner2000image}. Second, a segmentation is drawn for each voxel independently, following the categorical distribution specified by the (deformed) atlas at each location. And third, image intensities are drawn independently at each voxel, as independent samples of Gaussian mixture models conditioned on the underlying segmentation, i.e., each label has an associated set of Gaussian parameters describing the distribution of the intensities of its voxels.

In order to obtain an automated segmentation using the atlas, we first compute point estimates of the model parameters, namely the deformation of the atlas and the Gaussian parameters. This is done with a coordinate ascent strategy, in which the deformation and the Gaussian parameters are alternately updated while keeping the other fixed. The deformation is updated with a conjugate gradient algorithm, whereas the Gaussian parameters are estimated with the Expectation Maximization (EM) algorithm \citep{dempster1977maximum}. The deformation is initialized by fitting the atlas to the automated segmentation of brain structures provided by the FreeSurfer main \texttt{recon-all} stream (\emph{aseg.mgz}, \citealt{fischl2002whole}). Once the point estimates have been computed, the posterior probability of the segmentation is obtained as a by-product of the EM algorihtm. Further details are given in \citet{iglesias2015computational,puonti2016fast}.

An important design choice of the segmentation algorithm is which classes are grouped in tissue types. Forcing different labels with similar intensity characteristics (e.g., gray matter structures such as the cerebral cortex, the hippocampus and the amygdala) to share Gaussian parameters improves the robustness of the algorithm. Here, we chose to group the thalamic nuclei into three different sets, representing different tissue types. First, the reticular nucleus was grouped with the rest of cerebral white matter structures in the atlas, as the large amount of fibers that cross it gives it a very similar appearance to that of white matter. A second set includes the mediodorsal and pulvinar nuclei (i.e., MDm, MDl, PuA, PuM, PuL, and PuI). All other nuclei are grouped into a third set. The division of nuclei between the second and third sets reflects the internal boundary that can be observed in \emph{in vivo} MRI, even at standard resolution (see top row in Figure~\ref{fig:sampleSegs}). Fitting the atlas to this internal boundary provides a more reliable estimate of the segmentation.

%%%%%%%%%%%%%%%%%%%%%%%%%%%%%%%%%%%%%%%%%%%

\section*{Experiments and results}

\noindent
To validate the built atlas and its application to segmentation, we performed four different sets of experiments. As a first basic check, we conducted a volume comparison of the thalamic nuclei derived with the proposed atlas  with those obtained with the atlas described in \citealt{krauth2010mean} (also derived from histology) using registration-based segmentation. The other three experiments aimed at  evaluating the performance of the proposed segmentation method \emph{indirectly} -- as direct evaluation would require labeling \emph{in vivo} scans with the \emph{ex vivo} protocol, which is not feasible. Specifically, the second experiment evaluated the  reliability of our segmentation over time using an \emph{in vivo} test-retest T1 MRI dataset. The third set of experiments evaluated the robustness of the proposed atlas with respect to changes in MRI contrast of the input scan, using a heavily multimodal MRI dataset. The fourth and last set of experiments assessed the ability of the proposed method to detect differential effects in the thalamus in a group study of Alzheimer's disease, using the publicly available ADNI dataset. 

\subsection*{Volumetric comparison with Krauth's atlas}

\noindent
To initially examine our proposed atlas in relation to existing thalamic atlases derived from histology, we selected \citet{krauth2010mean}. We conducted volumetric analysis for six representative nuclei (one per thalamic group, see Table~\ref{tab:definitionOfNuclei}), which are also present in Krauth's atlas, and which are well characterized in terms of functional and structural connectivity: anteroventral (AV), lateral posterior (LP), centromedian (CM), mediodorsal (MD, equal to the union of MDl and MDm), ventral lateral (VL, equal to the union of VLa and VLp), and pulvinar (PU, equal to the union of PuA, PuM, PuL and PuI).

Comparing the volumes of the nuclei in the two atlases directly is problematic, particularly given that Krauth's model is a non-probabilistic mean derived from a small number of subjects. Instead, we compared the distributions of volumes of the nuclei when the atlases were applied to the automated segmentation of 66 subjects. For the proposed atlas, we used the method described above in this paper. For Krauth, we used direct registration-based segmentation: we took the MNI template with thalamic labels overlaid (see \citealt{jakab2012generation} for details); deformed it to the target subjects with ANTS \citep{avants2008symmetric}; and used the resulting deformation fields to warp the labels to target space and create the segmentations. 

The 66 subjects (age 24.31 $\pm$ 3.70 years; 40 females) were right-handed healthy adults, with no history of psychiatric, neurological, attention or learning disorders. They were scanned at the BCBL on a 3 T Siemens Magnetom TIM Trio scanner, using a 32-channel head coil. T1-weighted images were acquired with a ME-MPRAGE sequence \citep{van2008brain}  with TE = 1.64, 3.5, 5.36, 7.22 ms, TR = 2530 ms, TI=1100 ms, $\alpha = 7^\circ$, FOV = 256 mm, 176 slices, resolution 1 mm isotropic. 
% All of them gave written informed consent. The experiment was approved by the BCBL Ethics Review Board and complied with the guidelines of the Helsinki Declaration.  

\begin{figure}[t!]
\centering
\includegraphics[width=0.5\textwidth]{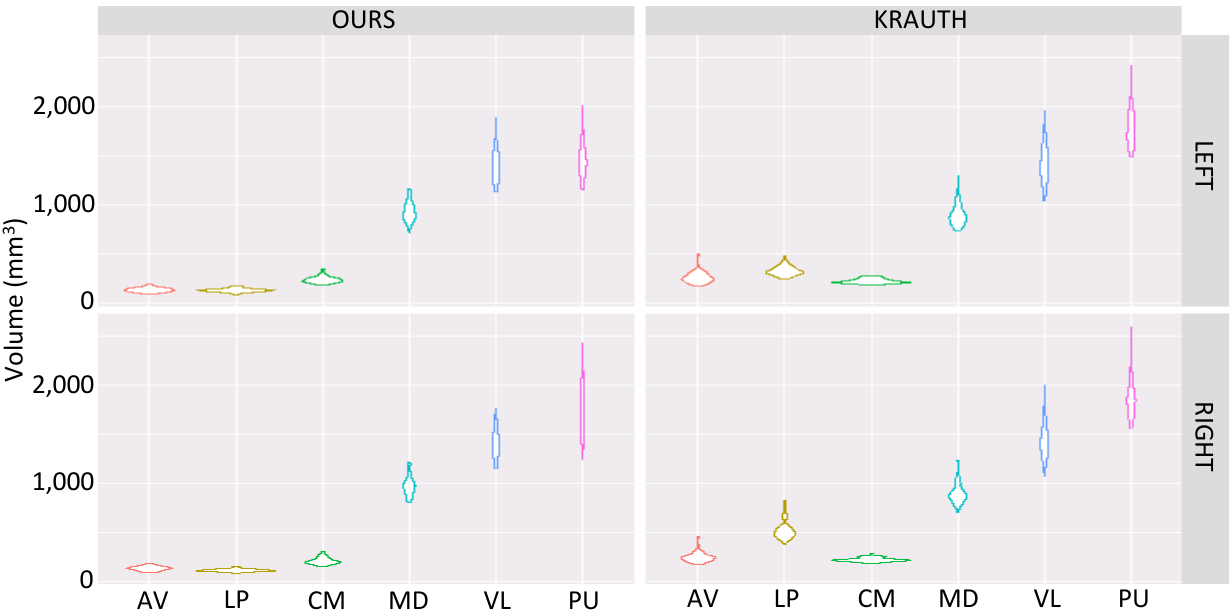}
\caption{Violin plots (box plots with a rotated kernel density plot on each side) for volumes of representative nuclei, computed with our proposed atlas and with Krauth's. MD, VL and PU represent the whole mediodorsal, ventral lateral and pulvinar nuclei, i.e., MD is the union of MDm and MDl;  VL is the union of VLa and VLp; and PU is the union of PuA, PuM, PuL and PuI.}
\label{fig:violin}
\end{figure}

\begin{figure}[t!]
\centering
\includegraphics[width=0.5\textwidth]{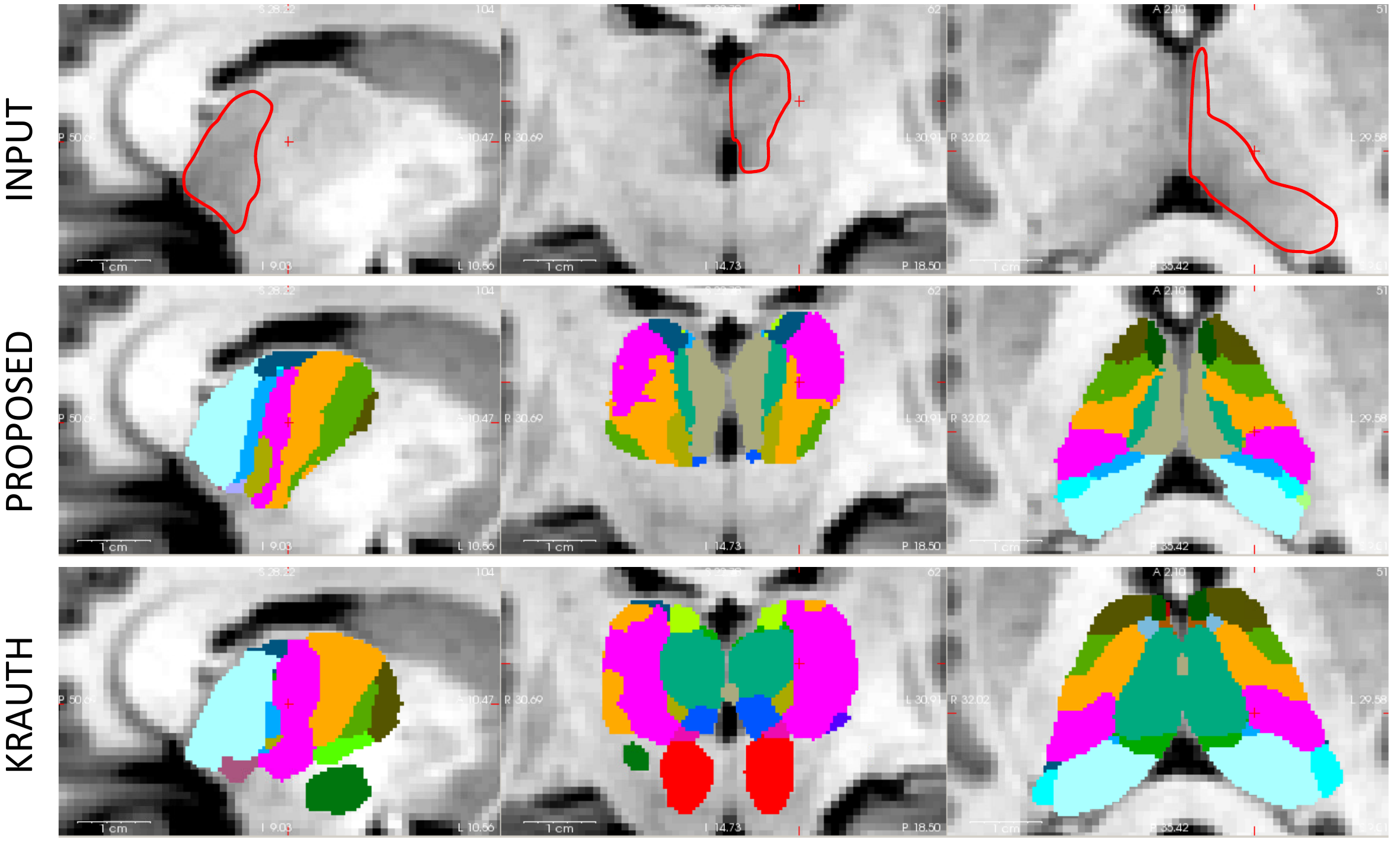}
\caption{Sagittal (left), coronal (middle), and axial (right) slices of a brain MRI scan segmented with the proposed atlas and Krauth's. The color map for our atlas is that described in Table~\ref{tab:definitionOfNuclei}. For Krauth, we attempted to match the color map as much as possible; we note that their atlas includes the red nucleus (in red) and subthalamic nucleus (in green), both in the inferior region, which are not part of our proposed atlas. On the left thalamus of the input scan, we have overlaid a manually delineated boundary between the  mediodorsal and pulvinar nuclei, and the rest of the nuclei; this is the main feature we use to fit the internal boundaries of the thalamus.  }
\label{fig:sampleSegs}
\end{figure}

Figure~\ref{fig:violin} displays violin plots (box plots with a rotated kernel density plot on each side) comparing the distribution of volumes of the representative nuclei. The agreement between the nuclei is high, even though Krauth's atlas yields slightly larger volumes for the anteroventral and lateral posterior nuclei (bilateral); and the left pulvinar nucleus. This is also apparent in Figure~\ref{fig:sampleSegs}, which displays a case segmented with both atlases. We note that direct comparison of whole thalamic volumes is not straightforward, due to different inclusion criteria, e.g., Krauth's atlas covers the red nucleus, subthalamic nucleus and mammillothalamic tract, while the proposed atlas does not (see Figure~\ref{fig:sampleSegs}). 
 
\subsection*{Test-retest reliability}

\noindent
In order to evaluate the test-retest reliability of the proposed segmentation method, we used MRI data from 31 of the 66 subjects (age 24.34 $\pm$ 2.96 years; 17 females), who also participated in a second session between seven and 10 days after the first session. The scanning protocol was the same as in the first experiment. The intraclass correlation coefficients (ICC) between the volumes derived from the scans in the first and second session are displayed in Table~\ref{tab:ICCs}. Despite the fact that the thalamus is notoriously difficult to segment due to its faint lateral boundary, our algorithm produces very high ICCs for the left and right whole thalami (above 0.97). Moreover, and despite the fact that the internal boundary used by the algorithm to fit the atlas is also faint, the ICCs for the individual representative nuclei are also excellent -- all scores are over 0.85, and most are over 0.90 and even 0.95.

\begin{table}[t]
\centering
\begin{tabular}{| l | l | l |}
\hline
Structure & Left & Right\\
\hline
Anteroventral (AV) & 0.86 & 0.93 \\ 
Lateral posterior (LP) & 0.85 & 0.90 \\ 
Centromedian (CM) & 0.94  & 0.92  \\ 
Mediodorsal (MD) & 0.89 & 0.95  \\ 
Ventral lateral (VL) & 0.96 & 0.99 \\ 
Pulvinar (PU) & 0.96 & 0.87 \\ 
\hline
Whole thalamus & 0.97 & 0.98 \\ 
\hline
\end{tabular}
\caption{Intraclass correlation coefficients for representative thalamic nuclei and whole thalamus. As in Figure~\ref{fig:violin},  MD is the union of MDm and MDl; VL is the union of VLa and VLp; and PU is the union of PuA, PuM, PuL and PuI.}
\label{tab:ICCs}
\end{table}

\subsection*{Robustness against changes in MRI contrast}

\noindent
Compared with discriminative approaches, a general advantage of Bayesian segmentation methods is their robustness against changes in the MRI contrast of the input scans. In order to test this robustness, we used a separate dataset consisting of multimodal MRI data from 11 subjects, acquired on a 3 T Siemens Prisma scanner (32 channel head coil) at the Martinos Center for Biomedical Imaging. For each subject, we first acquired two T1-weighted, MP2RAGE scans \citep{marques2010mp2rage} with parameters: TI = 700 ms and 2500 ms; $\alpha = 4^\circ$ and $5^\circ$, TE = 2.98 ms, image size = 256$\times$240$\times$176, GRAPPA accelaration factor = 3, bandwith 240Hz/pixel, resolution 1 mm isotropic. The two T1s were motion corrected and averaged. The two inversion times were used to compute the quantitative T1 relaxation time at each voxel. Given the quantitative T1 data and the individual MPRAGEs, we computed the proton density (PD). Using these data, we synthesized the k-space for a single-inversion MPRAGE with $\alpha = 7^\circ$, TR = 2530 ms, TE = 0 ms, with inversion times ranging from 300 ms to 940 ms using a Bloch simulator \citep{ma2013magnetic}. Within the simulation, the data were acquired instantaneously (i.e., infinite bandwidth). The simulated k-space data was then reconstructed using an FFT and the absolute value taken.

We also acquired two additional volumes for each subject. First, a T2-weighted volume with the following parameters: TR = 3200 ms, TE = 564 ms, acceleration factor = 2, bandwidth = 651 Hz/pixel, echo spacing 3.66 ms, resolution 1 mm isotropic. 
And second, an additional MPRAGE with a contrast that is typically used in DBS: TR = 3000 ms, TE = 3.56 ms, TI = 406 ms, $\alpha=8^\circ$, phase field of view = 81.3\%, resolution 0.8 mm isotropic. 

The segmentation algorithm for the alternative MRI contrasts is the same as for the T1 data, and also uses the automated segmentation \emph{aseg.mgz} in the initialization. Since this segmentation is computed by FreeSurfer from the T1 images, the results of this experiment are positively biased by the common initialization. However, we note that this will be the scenario of the public release of the algorithm in FreeSurfer, in which the availability of a T1 scan is always assumed.

\begin{figure}[t!]
\centering
\includegraphics[width=0.5\textwidth]{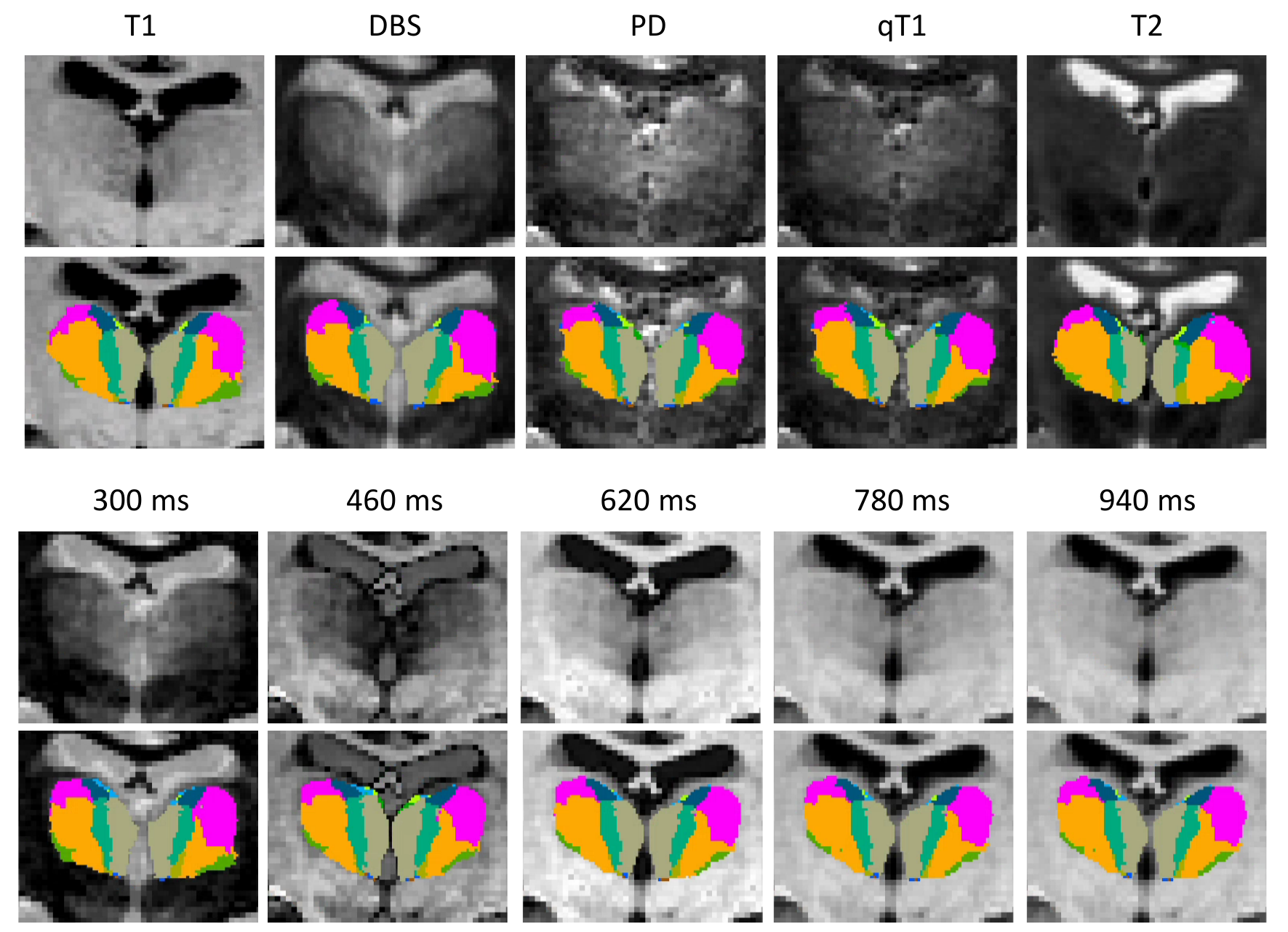}
\caption{Coronal slice from sample subject in multimodal MRI dataset, with corresponding automated segmentations with the proposed atlas.}
\label{fig:robustnessSamples}
\end{figure}

\begin{figure*}[t!]
\centering
\includegraphics[width=0.8\textwidth]{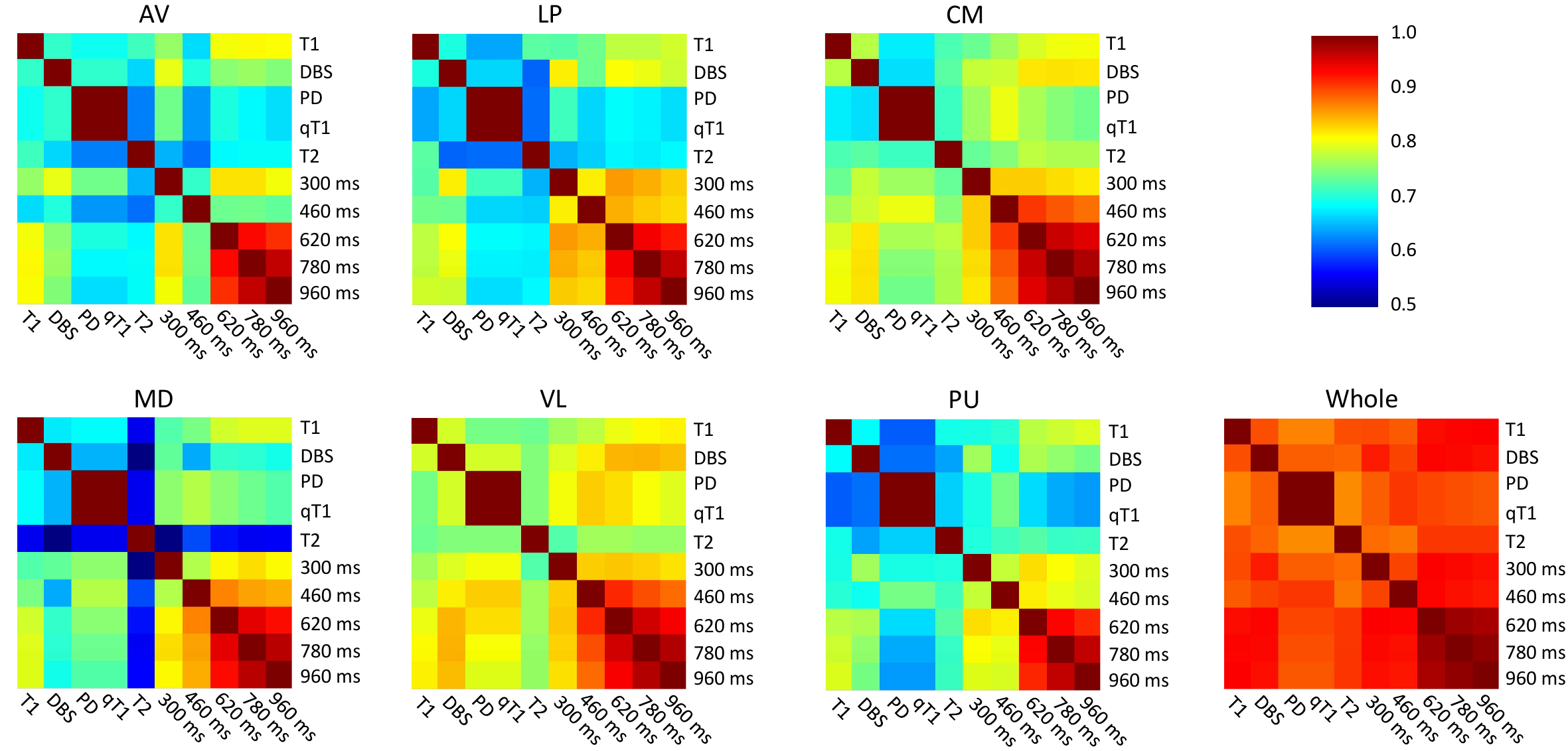}
\caption{Dice overlap (left-right averaged) for the segmentations yielded by different MRI contrasts. The six matrices correspond to the six representative thalamic nuclei and whole thalamus. As in previous figures, MD is the union of MDm and MDl; VL is the union of VLa and VLp; and PU is the union of PuA, PuM, PuL and PuI. The times in ms refer to the synthetic MPRAGE scans. The color bar is the same for all six matrices.}
\label{fig:robustness}
\end{figure*}

An example of a coronal slice with all available MRI contrasts and associated segmentations is shown in Figure~\ref{fig:robustnessSamples}. The segmentation is quite stable across contrasts, though some differences can be observed both in the internal and external boundaries of the thalamus. Quantitative results are displayed in Figure~\ref{fig:robustness}, which shows, for each representative nucleus, the Dice overlap between the segmentations for each pair of MRI contrasts. The agreement between the segmentations of the whole thalamus is quite high, near or above 0.90 in almost all cases. For the individual nuclei, the overlaps are moderately high, given that we are considering substructures, particularly for CM and VL (Dice approximately between 0.75 and 0.85). The overlaps are slightly lower for AV, MD and PU, with Dice scores approximately between 0.65 and 0.75. In terms of MRI contrast, the best agreement is observed (as expected) between the synthetic MPRAGEs with inversion times over 600 ms -- since TI $<$ 600 ms produces a contrast flip. The least consistent MRI contrast is the T2, in which the boundary between MD/PU and the rest of nuclei is almost invisible (see example in Figure~\ref{fig:robustnessSamples}). In general, the agreement is good between most pairs of contrasts.

\subsection*{Alzheimer's disease study}

\noindent
In order to assess the effectiveness in neuroimaging group studies of our proposed automated segmentation method based on the \emph{ex vivo} atlas, we ran the algorithm on a subset of the publicly available ADNI dataset.
The ADNI (\url{adni.loni.usc.edu}) was launched in 2003 as a public-private partnership, led by Principal Investigator Michael W. Weiner, MD. The primary goal of ADNI has been to test whether serial MRI, positron emission tomography, other biological markers, and clinical and neuropsychological assessment can be combined to measure the progression of mild cognitive impairment and early Alzheimer’s disease.
Here we considered T1-weighted scans from 213 subjects with Alzheimer's and 161 age-matched controls (Alzheimer's: 76.04$\pm$5.42 years; controls: 75.58$\pm$7.37 years); we note that these are the subjects that we have used in previous studies from our group (e.g., \citealt{iglesias2015computational,saygin2017high}). The resolution of the T1 scans was approximately 1 mm isotropic; further details on the acquisition can be found in the ADNI website. The volumes of the thalamic nuclei and of the whole thalamus were corrected by age and intracranial volume (as estimated by FreeSurfer) and left-right averaged in all analyses.

To discriminate the subjects into the two classes, we considered three different approaches. First, thresholding of the whole thalamic volume, as estimated with the main FreeSurfer \texttt{recon-all} stream (i.e, \emph{aseg.mgz}). Second, thresholding of the whole thalamic volume, as estimated by the proposed atlas (i.e., summing all the nuclei). And third, thresholding of the likelihood ratio given by a linear discriminant analysis (LDA, \citealt{fisher1936use}), computed in a leave-one-out fashion. LDA is a simple linear analysis, which enables us to consider all nuclei simultaneously, while ensuring that the performance is mostly determined by the input data, rather than stochastic variations in a more complex classifier.
 
\begin{figure}[t!]
\centering
\includegraphics[width=0.5\textwidth]{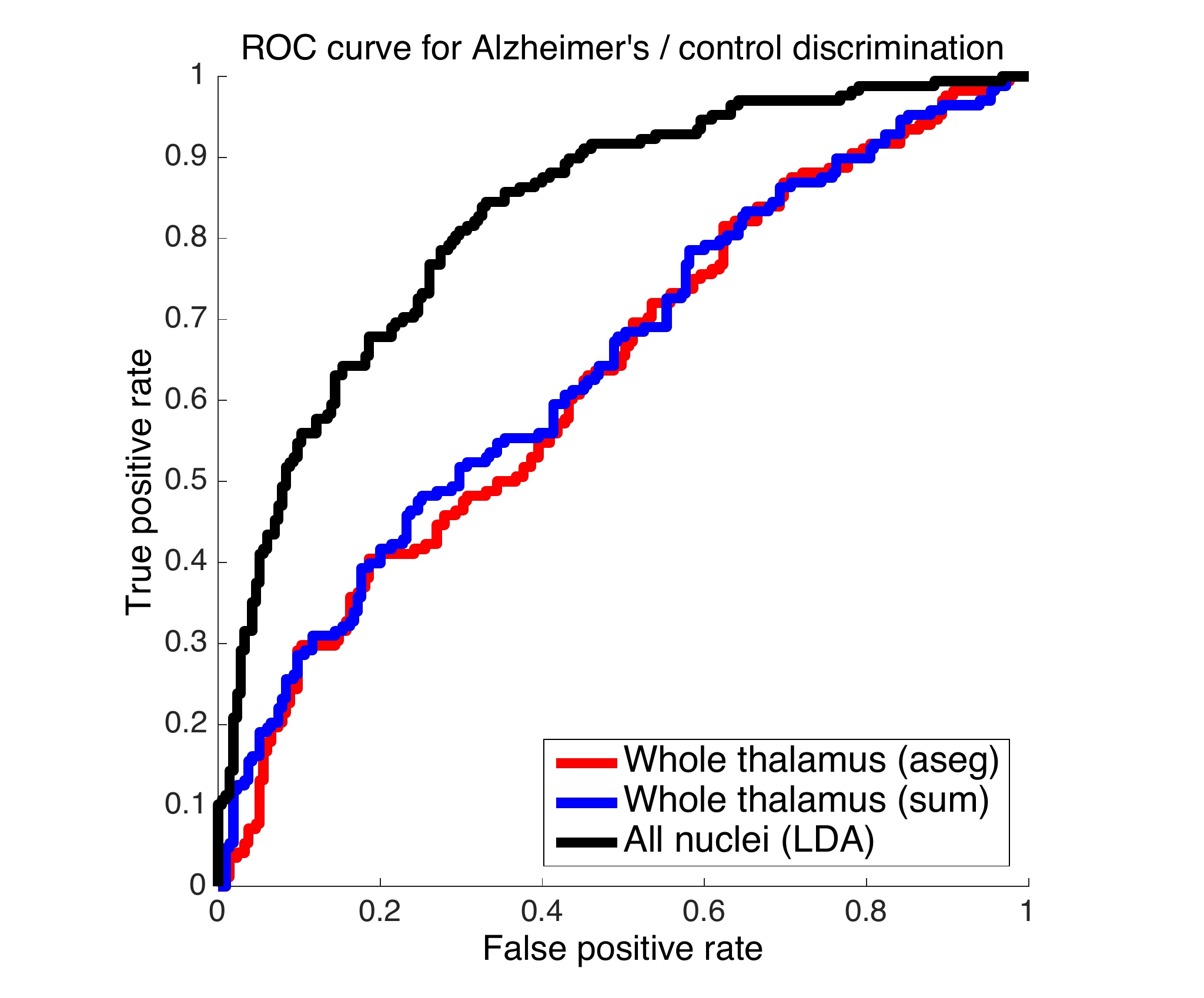}
\caption{ROC curves for Alzheimer's vs. controls classification based on left-right averaged thalamic volumes. The red curve is for the whole thalamic volume estimated by FreeSurfer's \texttt{recon-all} stream (AUC = 0.632). The blue curve is for the whole thalamic volume, estimated as the sum of the volumes of the nuclei given by our proposed atlas (AUC = 0.645). The black curve corresponds to a leave-one-out LDA classifier that simultaneously considers the volumes of all nuclei, as estimated by the proposed atlas  (AUC = 0.830). The volumes were left-right averaged and corrected by age and intracranial volumes in all cases.}
\label{fig:ROC}
\end{figure}

The receiver operating characteristic (ROC) curves are shown in Figure~\ref{fig:ROC}. The area under the curve (AUC), which is a threshold-independent measure of performance for a classifier, was 0.632 for the thalamic volumes given Freesurfer's main stream. When using the whole thalamic volumes given by the proposed  atlas, the AUC was slightly higher (AUC = 0.645), though the difference was not statistically significant ($p = 0.4$) according to a paired DeLong test \citep{delong1988comparing}. However, when using all nuclei simultaneously, the AUC increased considerably to 0.830 ($p \sim  10^{-10}$ against the whole thalamic volumes, given by \texttt{recon-all} or our proposed atlas).

\begin{table}[t]
\centering
\begin{tabular}{| l | l | l |}
\hline
Structure & Avg. vol. & p-value\\
\hline
Lateral geniculate (LGN) & 124  mm$^3$ & $ < 10^{-16}$  \\ 
Medial magnocellular (MDm) & 606 mm$^3$ & $ < 10^{-12}$ \\ 
Lateral parvocellular (MDl) & 245 mm$^3$&  $ < 10^{-9}$  \\ 
Medial  Geniculate (MGN) & 99  mm$^3$ & $ < 10^{-5}$  \\ 
Ventral anterior (VA) & 343 mm$^3$ & $ 0.0001$ \\ 
Anteroventral (AV) & 100 mm$^3$ & $ 0.0004$ \\ 
\hline
Whole thalamus & 5571 mm$^3$ & $0.0004$ \\ 
\hline
\end{tabular}
\caption{Thalamic nuclei showing statistically significant differences between Alzheimer's and controls, sorted by increasing p-value (two-side, non-parametric Wilcoxon test). The threshold for statistical significance is $p<0.0019$, which is equivalent to $p<0.05$ Bonferroni-corrected by 26 multiple comparisons. In all cases, the volume of the nuclei was bigger in the control population. The table also displays the average volume across the population, as a measure of the size of the nuclei.}
\label{tab:individualNuclei}
\end{table}

While AUC = 0.830 is modest in terms of Alzheimer's classification, it represents a very large increase with respect to using the volume of the whole thalamus alone. The reason for this increase is apparent from Table~\ref{tab:individualNuclei}, which shows the p-values for the individual nuclei that display significant differences between the two groups, after Bonferroni correction by the number of nuclei. The table shows that fitting the internal boundary of the thalamus, even if  faint (and hence prone to segmentation mistakes), enables some thalamic structures to separate the two classes with much more accuracy than the whole thalamus. These results are consistent with the literature \citep{aggleton2016thalamic,pini2016brain} and may be explained by the distinct pattern of connections and functions of these thalamic regions, and by the fact that not all thalamic nuclei are equally affected by the disease -- see further details below under Discussion.

\section*{Discussion}

\noindent
Human cortical information flow and dynamics cannot be fully understood without taking into account thalamocortical interactions. Due to its critical functions and widespread structural connections with practically the entire cortex, the availability of accurate and reliable automated segmentation algorithms for the thalamic nuclei is of high interest for the neuroimaging community. Here we have introduced a probabilistic atlas of 26 human thalamic nuclei built upon 3D reconstructed histological data from 12 thalami; presented a Bayesian segmentation method that applies the atlas to the automated segmentation of thalamic nuclei in \emph{in vivo} brain MRI scans of arbitrary contrast; and validated the atlas and segmentation with four different sets of experiments, whose results are discussed next. 

First, we compared the proposed atlas with a previously presented histology-based atlas \citep{krauth2010mean} by segmenting the thalamic nuclei in a population of 66 subjects and comparing the distribution of the volumes. To produce the segmentations, our proposed atlas was combined with a Bayesian segmentation technique presented in this article. For Krauth, we used direct registration-based segmentation. Leaving aside discrepancies in the anatomical definition of nuclei, the two models yielded very similar volumes, which is reassuring in terms of scientific reproducibility. The advantage of our model lies in its probabilistic nature, which enables segmentation of scans of arbitrary MRI contrast withing a Bayesian framework.

Second, we conducted a test-retest reliability study of our segmentation tool using 1 mm T1 scans acquired approximately one week apart. This experiment revealed excellent repeatability, with ICC scores mostly above 0.90  (which is higher than that of some whole structures in FreeSurfer; see \citealt{morey2010scan}). This result reassures that the obtained volumes are reliable, i.e., they are an accurate representation of a measurement, rather than attributed to random fluctuations. 

A third experiment assessed the  robustness of the tool to changes in the MRI contrast of the input scan. This experiment was also successful, as the agreement between segmentations is good across a wide array of contrasts. This result supports the hypothesis that our structural segmentation algorithm will be able to take advantage of MRI pulse sequences that produce good contrast in the thalamus \emph{in vivo}, such as fast gray matter acquisition T1 inversion recovery scans (FGATIR).  

The most intriguing results are those from the experiment with Alzheimer's disease subjects, as a large boost in classification performance (increase of 0.20 in AUC) is observed when switching from the volume of the whole thalamus to the volumes of its subregions. Looking at individual nuclei, large differences were found in mediodorsal, anteroventral and ventral anterior areas. This is consistent with previous results in neuroimaging studies. Antero-ventro-medial regions have been reported as main foci of impairment in Alzheimer's disease in several neuroimaging studies \citep{aggleton2016thalamic,pini2016brain,stepan2014cortical,zarei2010combining}. Moreover, a three-year longitudinal study reported that thalamic atrophy was first localized in the ventromedial  regions, and spread to anterior regions in later stages \citep{cho2014shape}, which are well represented in the ADNI dataset. 
Mediodorsal regions have also been reported as foci of atrophy: for example, a voxel-based morphometry study in genetic Alzheimer's disease showed that mutation carriers exhibit a decreased gray matter density localized in the medial-dorsal regions of the thalamus within 5 years of symptoms onset \citep{cash2013pattern}.

Our results are also supported by neuropathological studies. For example, \citet{braak1991alzheimer} and \citet{xuereb1991nerve} have shown that the primary site of Alzheimer's disease degeneration in the thalamus is the anterodorsal nucleus, which showed severe neuronal loss and tangle formation. In our proposed atlas,
% based on Jones's characterization of the mammalian thalamus \citep{jones2012thalamus}, 
the anterodorsal nucleus (due to its small size) has been included in the anteroventral (AV) nucleus. The AV nucleus not only shows a strong effect between Alzheimer's and controls, but is also the nucleus with the strongest ICC in the repeatability experiment. 
These anterior thalamic nuclei are densely and directly connected to the medial temporal lobe structures (e.g., hippocampus, amygdala and entorhinal cortex), in addition to the relay in the mammillary nuclei of the hypothalamus, all of which are  typically affected in Alzheimer's disease and linked to its episodic memory deficits \citep{aggleton2016thalamic}. 
Moreover, \citet{zarei2010combining} showed that atrophy of the mediodorsal thalamus (which shows the second to largest effect in our experiment) corresponds to changes in connectivity with cortical and subcortical areas. % the anterior temporal cortex and the posterior hippocampus. 
Finally, \citet{braak1991alzheimer} also found amyloid deposition in the anteroventral, laterodorsal, and the central medial nucleus; the first two of these regions (anteroventral and laterodorsal) correspond with our volumetric findings. 

While these neuroimaging and neuropathological studies support the outcome of our experiment, our results could also be a partial correlate of the expansion of the neighboring ventricles in Alzheimer's disease. For example, the thalamic atrophy detected by \citealt{zarei2010combining} through a shape analysis was characterized by an inward movement of vertices in the dorsomedial and ventral aspects of the thalamus, which may be associated with ventricular enlargement. However, we may also argue that such ventricular effect should also affect the volume of the whole thalamus, which is only the case to a much lesser extent in our experiment. Another aspect that requires further inspection is the atrophy detected in the lateral and medial geniculate nuclei (LGN/MGN), for which there is little evidence in the literature. The measured atrophy might be a false positive due to small nuclei sizes, lack of contrast with neighboring cerebral white matter,
% and shift of the posterior horn of the lateral ventricles. 
or both.
However, it could also be a true effect, e.g., a correlate of the visual impairments associated with Alzheimer's disease \citep{kirby2010visual}, as the LGN is a major relay of the visual pathway; a similar effect is potentially possible for the MGN, which is linked to auditory processing. We note that, even when the LGN (the single most discriminating nucleus; see Table~\ref{tab:individualNuclei}) and the MGN are removed from the LDA analysis, the AUC of the classifier is still 0.783, which is still considerably higher than the values given by the whole thalamus (AUC = 0.645).

\section*{Conclusion}

\noindent
We have presented a probabilistic atlas of the human thalamus based on \emph{ex vivo} imaging techniques (\emph{ex vivo} MRI, histology), and a companion  tool that enables segmentation of thalamic nuclei from \emph{in vivo} MRI of arbitrary contrast. At the technical level, future work will focus on integrating diffusion MRI data in the segmentation algorithm. While segmentation based solely on structural MRI enables analysis of large amounts of legacy datasets that do not include diffusion data, the local diffusion information and structural connectivity are strong signatures of the divisions between different thalamic nuclei. Therefore, including these data in the generative model should greatly inform the Bayesian segmentation algorithm, which currently relies on faint boundaries and prior knowledge to fit the atlas to the input images. 

Improvements in the segmentation will also enable more advanced studies of multiple disorders, e.g., Alzheimer's disease, Parkinson's disease, dyslexia, schizophrenia, etc. Future analyses will include: further investigation of the results of the group study reported in this article; correlation of thalamic nuclei with clinical scores, subtypes of the disease, and disease duration; investigation of specific thalamic networks (functional and structural) using the segmented nuclei as seeds; or cluster analysis of nuclei volumes for subtyping the disease. 

Our proposed atlas and segmentation tool are publicly available as part of the neuroimaging software package FreeSurfer (\url{https://surfer.nmr.mgh.harvard.edu}), and will enable neuroimaging studies of the human thalamus at sites that do not possess the expertise or staff resources to manually delineate the thalamic nuclei in 3D MRI data.  

\section*{Acknowledgements}

\noindent
The authors would like to thank Professor Karla Miller (Oxford) for her help with the design of the \emph{ex vivo} MRI acquisition; Ms. Mercedes \'I\~{n}iguez de Onzo\~{n}o and Mr. Francisco Romero (UCLM) for their careful technical laboratory help; and Mr. Gonzalo Artacho (UCLM) for his help with the digitization and curation of his organization of histological data.

% Eugenio 
This project has received funding from the European Union's Horizon 2020 research and innovation program under the Marie Sklodowska-Curie grant agreement No 654911 (project ``THALAMODEL'') and by the ERC Starting Grant agreemnent No 677697 (``BUNGEE-TOOLS''). It was also funded by the Spanish Ministry of Economy and Competitiveness (MINECO TEC-2014-51882-P, RYC-2014-15440, and PSI2015-65696), the Basque Government (PI2016-12), and UCLM Internal Research Groups grants.

Support for this research was also provided in part by the National Institute for Biomedical
Imaging and Bioengineering (P41EB015896, 1R01EB023281,
R01EB006758, R21EB018907,
R01EB019956), the National Institute on Aging (5R01AG008122,
R01AG016495), the National Institute of Diabetes and Digestive and
Kidney Diseases (1-R21-DK-108277-01), the National Institute for
Neurological Disorders and Stroke (R01NS0525851, R21NS072652,
R01NS070963, R01NS083534, 5U01NS086625), and was made possible by the
resources provided by Shared Instrumentation Grants 1S10RR023401,
1S10RR019307, and 1S-10RR023043. Additional support was provided by the NIH
Blueprint for Neuroscience Research (5U01-MH093765), part of the
multi-institutional Human Connectome Project. In addition, B.F. has a
financial interest in CorticoMetrics, a company whose medical pursuits
focus on brain imaging and measurement technologies. B.F.'s interests were
reviewed and are managed by Massachusetts General Hospital and Partners
HealthCare in accordance with their conflict of interest policies.

% ADNI blob
Data collection and sharing for this project was funded by the Alzheimer's Disease Neuroimaging Initiative (ADNI) (National Institutes of Health Grant U01 AG024904) and DOD ADNI (DoD award number W81XWH-12-2-0012). ADNI is funded by the National Institute on Aging, the National Institute of
Biomedical Imaging and Bioengineering, and through generous contributions from the following: AbbVie, Alzheimer’s Association; Alzheimer’s Drug Discovery Foundation; Araclon Biotech; BioClinica, Inc.; Biogen; Bristol-Myers Squibb Company; CereSpir, Inc.; Cogstate; Eisai Inc.; Elan Pharmaceuticals, Inc.; Eli Lilly and Company; EuroImmun; F. Hoffmann-La Roche Ltd and its affiliated company Genentech, Inc.; Fujirebio; GE Healthcare; IXICO Ltd.; Janssen Alzheimer Immunotherapy Research \& Development, LLC.; Johnson \& Johnson Pharmaceutical Research \& Development LLC.; Lumosity; Lundbeck; Merck \& Co., Inc.; Meso Scale Diagnostics, LLC.; NeuroRx Research; Neurotrack Technologies; Novartis Pharmaceuticals Corporation; Pfizer Inc.; Piramal Imaging; Servier; Takeda Pharmaceutical Company; and Transition Therapeutics. The Canadian Institutes of Health Research is providing funds to support ADNI clinical sites in Canada. Private sector contributions are facilitated by the Foundation for the National Institutes of Health (www.fnih.org). The grantee organization is the Northern California Institute for Research and Education, and the study is coordinated by the Alzheimer’s Therapeutic Research Institute at the University of Southern California. ADNI data are disseminated by the Laboratory for Neuro Imaging at the University of Southern California.

\bibliographystyle{model2-names}
\bibliography{mybibfile}

\end{document}